# Precision Measurements of Ice Crystal Growth Rates

Kenneth G. Libbrecht[1]
Department of Physics, California Institute of Technology
Pasadena, California 91125

**ABSTRACT**
I describe techniques for making precise measurements of the growth rates of the principal facets of ice crystals. Particular attention is paid to identifying and reducing systematic errors in the measurements, as these have plagued earlier attempts to determine ice growth rates. I describe the details of an experimental apparatus we are currently using, and I describe some preliminary results for growth of basal facets at $T = -15$ C.

[This paper is also available (with better formatting and perhaps corrections) at http://www.its.caltech.edu/~atomic/publist/kglpub.htm.]

## 1. Introduction

The growth of snow crystals from water vapor in air is governed by a number of factors, with vapor diffusion and attachment kinetics at the ice surface being the dominant players. While vapor diffusion is well known and calculable in principle, our understanding of the attachment kinetics controlling ice crystal growth remains quite incomplete. As a result, many observations of the morphology of ice crystals grown under different conditions remain unexplained [1]. In particular, the growth morphology of snow crystals is known to change dramatically with temperature over the range $-30$ C $< T < 0$ C, and at present there is not even a satisfactory qualitative explanation of this growth behavior.

Accurate measurements of the growth rates of ice crystals under different conditions are necessary for constraining models of the growth process, and thus for investigating attachment kinetics. In a careful analysis of past experiments, however, we have found that existing growth data are largely unreliable [2]. Systematic errors of various types affected the measurements in substantial ways, and it now appears that these effects were not properly dealt with in any previous experiments [2]. The goal of the present paper is to identify and investigate these systematic effects in a quantitative fashion, and to describe an experimental apparatus and procedure that is capable of making accurate measurements of ice crystal growth rates under a variety of conditions.

## 2. Notation and Measurement Strategy

Following the notation of [1], we write the growth velocity normal to the surface in terms of the Hertz-Knudsen formula

$$v_n = \alpha \frac{c_{sat}}{c_{solid}} \sqrt{\frac{kT}{2\pi m}} \sigma_{surf} \qquad (1)$$
$$= \alpha v_{kin} \sigma_{surf}$$

---

[1]e-mail address: kgl@caltech.edu

where the latter defines the velocity $v_{kin}$. In this expression $kT$ is Boltzmann's constant times temperature, $m$ is the mass of a water molecule, $c_{solid} = \rho_{ice}/m$ is the number density for ice, $\sigma_{surf} = (c_{surf} - c_{sat})/c_{sat}$ is the supersaturation just above the growing surface, $c_{surf}$ is the water vapor number density at the surface, and $c_{sat}(T)$ is the equilibrium number density above a flat ice surface. Experiments with ice growing from vapor are nearly always in a near-equilibrium regime, where $\sigma_{surf} \ll 1$.

The parameter $\alpha$ is known as the *condensation coefficient*, and it embodies the surface physics that governs how water molecules are incorporated into the ice lattice, collectively known as the *attachment kinetics*. The attachment kinetics can be nontrivial, so in general $\alpha$ will depend on $T$, $\sigma_{surf}$, and perhaps on the surface structure and geometry, surface chemistry, and other factors. If molecules striking the surface are instantly incorporated into it, then $\alpha = 1$; otherwise we must have $\alpha < 1$. The appearance of crystal facets indicates that the growth is limited by attachment kinetics, so we must have $\alpha < 1$ on faceted surfaces. For a molecularly rough surface, or for a liquid surface, we expect $\alpha \approx 1$.

Expressing the growth velocity in terms of an attachment coefficient carries with it an implicit assumption that the growth dynamics is effectively local in character [1]. Diffusion of water molecules along the ice surface, and especially between different facets, may mean this assumption is incorrect. There are theoretical reasons for believing that surface diffusion around corners is negligible in ice growth [1], so for the remainder of this discussion we will be assuming that the growth dynamics can be adequately expressed in terms of a condensation coefficient $\alpha(T, \sigma_{surf})$ that depends only on temperature and supersaturation at the growing surface.

Our measurement strategy will be quite simple – measure the growth velocity $v_n$ of a given crystal surface, determine $\sigma_{surf}$ from other measurements in the experiment, and determine $\alpha$ from Equation 1. Figure 1 shows the basic geometry of our measurements. We place a single, faceted ice crystal on a temperature-controlled substrate surrounded by an ice reservoir, where both the sample crystal and reservoir are inside a vacuum chamber with most of the air removed. We lower $\Delta T = T_{sample} - T_{reservoir}$, the temperature of the ice sample relative to the ice reservoir, causing the crystal to grow. We then measure crystal growth along the substrate with simple imaging, using a microscope objective just below the crystal and an external camera. We measure growth perpendicular to the substrate using laser interferometry. We shine a low-power Helium-Neon laser up through the microscope objective (see Figure 1), where it focuses to a several-micron spot on the crystal. The laser spot is reflected both from the substrate/ice interface and the ice/vacuum interface. The indices of refraction are such that the two reflected beams have roughly equal amplitude, so they interfere with one another. As the crystal grows, the imaged spot cycles in brightness, with a complete cycle taking place when the crystal thickness changes by $\Delta t = \lambda_{laser}/2n_{ice} = 243$ nm. With modest effort, the relative accuracy of this interferometric measurement is sufficient to observe thickness changes of less than a single water monolayer. (At the temperatures we operate, however, we expect that single-molecule steps are not sufficiently stable for imaging over millisecond timescales.) The camera image shows the outline of the crystal along with the interference spot from the laser, as shown in Figure 2.

We determine $\sigma_{surf}$ from $\Delta T$, and we determine $v_n$ from counting laser fringes (bright/dark cycles) with time, and we then calculate $\alpha(T, \sigma_{surf})$ using Equation 1. Once we have measured $\alpha(T, \sigma_{surf})$ over a range of temperatures and supersaturations, we can compare with crystal growth models to extract various growth parameters.

## 3. Experimental Apparatus

To make the necessary sample crystals for our measurements, we use a large, air-filled *nucleation chamber* shown schematically in Figure 3. For clarity, not all parts are drawn to scale. This chamber is approximately 90 cm tall and 50x50 cm in cross-section, and is constructed from 3-mm-inch thick copper plates to which copper cooling pipes have been soldered. We run chilled methanol through the pipes to cool the chamber to our desired temperature. For measuring basal growth rates (as in Figure 2), we typically run the large chamber near -15 C, where plate-like crystals grow.

The air in the nucleation chamber is supersaturated by evaporation from a heated vessel of water near the bottom of the tank. Convection carries water vapor up where it mixes with the air. In steady state, there is a flow of water vapor from the supply vessel to the air to the chamber walls, and this flow keeps the air inside the chamber supersaturated. The water vapor may also condense into water droplets above the vessel, and the movement of these supercooled droplets through the chamber also supersaturates the air. At temperatures as low as -15 C, not many ice crystals form in the air without the application of some nucleation agent.

We have not done extensive measurements of the supersaturation inside the chamber, but we expect it is quite variable, going to zero near the walls (once they become covered with frost, which happens fairly quickly). The supersaturation increases as more power is sent to the water heater, but we expect that the formation of droplets in the chamber limits the supersaturation to values not much above that for supercooled water droplets. We typically run the heater at about 15 Watts, which raises the air temperature in the chamber a few degrees above the temperature of the walls.

We use a rapid release of compressed air to nucleate the growth of ice crystals in this chamber (see Figure 3). Compressed nitrogen gas at 40-60 psi is fed into a U-tube inside the chamber, which is terminated with two valves. A small ice reservoir at the bottom of the U-tube (not shown in Figure 3) serves to saturate the gas at the ambient temperature. To make crystals, the first valve is opened to fill the space between the valves with gas, after which it is closed and the second valve is open. The sudden decompression cools the gas so that homogeneous nucleation generates small ice crystals [3].

The freshly nucleated crystals float and grow in the supersaturated air inside the chamber, until they become large enough to fall from gravity. (Convection also carries the crystals throughout the chamber.) It typically takes 1-5 minutes for the growing ice crystals to fall to the bottom of the chamber. To produce a steady flux of new crystals, we found it helped considerably to run the nucleator continuously from a timer, so the valves cycled approximately every 20 seconds. With the continuous cycling of the valves, along with the continuous evaporation from the water vessel, we achieved a steady density of small crystals growing and falling inside the chamber.

To move ice crystals from the nucleation chamber to the substrate inside the growth chamber, we slowly draw air from the inner chamber using a vacuum pump

(not shown in Figure 3). With a partial vacuum in the growth chamber, we open a valve to the larger chamber (V1 in Figure 3; V2 remains open at this time; its role will be discussed later). As air rushes into the growth chamber, it brings some ice crystals with it. By chance, occasionally an ice crystal lands on the substrate, where its growth can be measured. If a given pulse of air does not yield a satisfactory crystal, the procedure is repeated. The substrate is kept slightly warmer than the rest of the growth chamber during this process, so any unsatisfactory crystals will evaporate away. Once a suitable crystal falls on the substrate, its temperature is quickly reduced to stabilize the crystal.

We chose a sample crystal for study if it is visually in good shape (i.e., its morphology is that of a well-formed ice prism) and its overall size is between 20 and 50 microns. We also require that only a single sample crystal is present on the substrate. We often achieve this end by vaporizing neighboring crystals using the $CO_2$ laser shown in Figure 3, as is discussed below.

We have found that the growth chamber temperature must be within a few degrees of the nucleation chamber during the crystal transfer process. If not, the small crystals do not survive the rapid change in conditions, particularly the rapid change is the partial pressure of water vapor. Once a crystal has been transferred and stabilized, however, we can then slowly change the temperature of the growth chamber to reach a desired running temperature. It typically takes a few minutes to acquire a suitable crystal on the substrate, and it may take an additional 5-10 minutes to change the growth chamber temperature before the measurements can commence. In the end, we are able to obtain measurements on perhaps 1-2 crystals per hour of run time.

Figure 4 shows a schematic diagram of the growth chamber in our apparatus, where again, for clarity, not all parts have been drawn to scale. The chamber is cylindrically symmetrical with an inner diameter of 7 cm and an inner height of 2 cm. In the next section we will discuss a number of design features of this chamber.

## 4. Systematic Errors in the Measurements

Although our basic measurement strategy is quite simple, experience dictates that we must be exceedingly careful to examine systematic errors that can affect the measurements. We found that it was necessary to identify and reduce many of these systematic effects before we were able to obtain satisfactory data. Knowing the supersaturation, $\sigma_{surf}$, just above the surface of our growing ice crystal is the most challenging part of this experiment, and we found it necessary to put considerable effort into controlling the supersaturation field $\sigma(x)$ inside the growth chamber. In this section we discuss some potential systematic effects that can corrupt ice crystal growth measurements.

**Diffusion-Limited Growth.** In the presence of a background gas, the supersaturation near the surface of a growing crystal, $\sigma_{surf}$, is lower than the supersaturation of the surroundings. Indeed, a gradient in the supersaturation field is necessary for providing the flux of water vapor for the growing crystal. Understanding how diffusion affects $\sigma_{surf}$ is an important consideration in ice crystal growth measurements. The ramifications of this systematic effect have often been underestimated in previous experiments [2].

Following [1], the spherically symmetric case is instructive for looking at how diffusion affects the measured growth rates. In this case we can write the growth velocity as

$$v_n = \frac{\alpha \alpha_{diff}}{\alpha + \alpha_{diff}} v_{kin} \sigma_\infty \qquad (2)$$

$$= \frac{\alpha}{\alpha + \alpha_{diff}} \frac{c_{sat} D \sigma_\infty}{c_{solid} R}$$

where

$$\alpha_{diff} = \frac{c_{sat} D}{c_{solid} v_{kin} R} = \frac{D}{R} \sqrt{\frac{2\pi m}{kT}}, \qquad (3)$$

$v_{kin}$ was defined in Equation 1, $\sigma_\infty$ is the supersaturation far from the growing crystal, and $R$ is the sphere radius. For the specific case of ice growing at $T = -15$ C in air we have

$$\alpha_{diff}(-15C) \approx 0.15 \left(\frac{1\,\mu\text{m}}{R}\right)\left(\frac{D}{D_{air}}\right) \qquad (4)$$

where $D_{air} \approx 2 \times 10^{-5}$ m$^2$/sec is the diffusion constant for water vapor in air at a pressure of one atmosphere.

Diffusion has a negligible effect on the growth when $\alpha \ll \alpha_{diff}$, which can be achieved by reducing the air pressure in the growth chamber, since to lowest order $D \sim P^{-1}$. Diffusion effects are also reduced by making measurements using smaller crystals. For our typical growth measurements at $T = -15$ C, with crystals of order 40 $\mu$m in size and $P \approx 3$ Torr, we have $\alpha_{diff} \approx 0.9$, which is substantially greater than a typical measured $\alpha$. Thus, the effects of diffusion on our results are acceptably small, especially at small $\sigma_{surf}$ when $\alpha$ is especially small. Diffusion effects are not completely insignificant, however, and they become worse at higher temperatures.

For nonspherical crystals, the diffusion corrections become greater when one is measuring a slow-growing facet next to fast growing facets. For example, referring to Figure 2, we may encounter a situation when the basal facet is growing slowly while the lateral growth of the prism facets is much faster because of substrate interactions (discussed below). In such circumstances, the above analysis will likely underestimate the diffusion effects.

The prism case can be analyzed in more depth using numerical solutions to the diffusion equation, and these can be made substantially simpler by assuming a cylindrically symmetrical crystal shape [1]. We have examined several cases corresponding to actual data taken with plate-like crystals growing at -15 C, where all the crystal dimensions and growth velocities are known. We found that a correction of the data for diffusion may reduce our estimate of $\sigma_{surf}$ by as much as 20 percent while increasing our estimate of $\alpha$ by a similar amount. Both corrections show up in a plot of $\alpha(\sigma)$. In what follows, we have ignored these corrections to the data. As the scatter in our measurements diminishes, these diffusion effects will become more pronounced and will need to be examined with additional care.

**Neighboring Crystals.** The above diffusion analysis assumes a single, isolated crystal growing on the substrate. Our transfer process, however, rarely yields one crystal with a simple prism morphology. Neighboring crystals act as water vapor sinks that reduce $\sigma_{surf}$ near the sample crystal. A good way to think about the effects of neighboring crystals is to consider the supersaturation field $\sigma(x)$ inside the growth chamber. This field must satisfy Laplace's equation with the appropriate boundary conditions [1]. A growing crystal reduces the supersaturation near its surface, and this affects $\sigma(x)$ elsewhere because it changes the boundary conditions.

Modeling the effects of neighbor crystals is exceedingly difficult, in part because they change in size and shape as both they and the sample crystal grow. Nonfaceted neighbors are particularly efficient water vapor sinks, so these produce greater changes in the supersaturation profile than do faceted crystals. We have looked extensively at the effects of neighboring crystals in our growth chamber, both seen and unseen, and have found that it is essential to only consider data where the substrate contains a single growing crystal. A few neighbor crystals can easily reduce the growth velocities by a factor of three or more. Furthermore, it is important that all unseen crystals lie on surfaces that have the same temperature as the ice reservoir.

We only began obtaining consistent growth data when we added an additional laser (shown in Figure 3) to remove unwanted crystals from our substrate. We used a $CO_2$ laser for this purpose because the absorption depth of 10-micron light in ice is only a few microns. Thus a focused laser will tend to vaporize the unwanted crystals and not the crystal of interest. Some laser light also strikes and heats the substrate, but with care we have been able to evaporate neighboring crystals without seriously damaging our sample crystals.

When transferring crystals, we typically keep the substrate temperature slightly above the chamber temperature, so that transferred crystals evaporate away in about one minute. Then we draw some air out of the chamber, open V1 quickly to get a pulse of air from the large tank, and look to see what falls on the substrate. If we see nothing, or if the crystals have non-ideal morphologies, we let them evaporate away before trying another pulse. Many crystals in the nucleation chamber are poorly formed, so we often have to try many pulses. Eventually we obtain a crystal that has a clean hexagonal prism morphology without too many neighbors. Then we quickly lower the substrate temperature to obtain $\sigma_{surf} \approx 0,$ so the crystal is neither growing nor evaporating appreciably. We then use the $CO_2$ laser to carefully burn away the neighbors, and if all goes well we are left with a single, well-formed prism crystal lying flat on the substrate, with no neighbors, so we can commence with a growth run.

**Substrate Interactions.** The growth of a facet surface is typically limited by 2D nucleation on the surface, so any extraneous source of atomic steps may increase the growth rates. If a facet plane intersects a substrate, then interactions with the substrate may provide a source of steps, thus increasing the crystal growth rate. This phenomenon has been seen in other experiments [2], and it is an important systematic effect in ice growth measurements.

Figure 5 shows an example of a case when substrate interactions increased the crystal growth rates. We avoid this systematic error by measuring growth perpendicular to the substrate, as described in the experimental section.

**Temperature Gradients in the Growth Chamber.** When air is drawn into the growth chamber from the large tank, ice crystals are drawn in as well, and these can deposit all over the inside walls of the growth chamber. All these crystals, in addition to the ice reservoir, are then sources of water vapor for the sample crystal growing on the substrate. Since $\sigma_{surf}$ is determined from $\Delta T = T_{sample} - T_{reservoir}$, we must ensure that the entire growth chamber is at a uniform temperature (except for the substrate).

To this end, we constructed the growth chamber with copper walls no thinner that 0.25 inches, and insulated it from its surroundings. The ZnSe lens (see Figure 4) cannot be fully insulated from its surroundings (since light must pass through it), so we made this optical element from two thin lenses separated by an air gap. ZnSe is a good thermal conductor, and we made some effort to reduce any extraneous heat load on this lens stack.

Our thermal modeling of the chamber suggests that any temperature differential across the chamber is no more than 0.01 C. This equates to a relative error in the supersaturation of

$$\frac{\delta\sigma}{\sigma} = \frac{d\sigma}{dT}\frac{\delta T}{\sigma}$$

where $\delta T$ is the effective error in $\Delta T$ used to calculate $\sigma_{surf}$. For growth temperatures near −15 C, this gives $\delta\sigma/\sigma$ of less than 15 percent, which is acceptable. The correction becomes worse at higher temperatures when typical values of $\sigma$ become smaller.

This problem is probably not as bad as it seems for two reasons. First, the contribution to $\sigma_{surf}$ comes from all parts of the growth chamber, and especially the ice reservoir, which has the greatest ice mass and surface area. Since the temperature variations are highest at the extremities of the chamber (for example on the ZnSe lens), which have little ice mass, these contribute only a small amount $\sigma_{surf}$. Second, we normalize to $\sigma_{surf} = 0$ by increasing the temperature of the sample crystal until it just begins to evaporate. Even if there are some temperature gradients in the chamber, this normalization will reduce their effects as long as the gradients (and ice masses) remain constant during the short duration of a growth measurement. In our calibration step, we are able to determine the relative temperature for the onset of evaporation to about 5 millidegrees, which corresponds to a supersaturation error of $\delta\sigma \approx 5 \times 10^{-4}$.

**Errors from Pumping on the Chamber**. Once a sample crystal has been successfully transferred to the substrate, we pump the air out of the chamber to produce a near vacuum inside. To reduce any problems associated with the pumpout tube, we made the section connecting to the growth chamber out of a 20-cm long section of copper pipe with a diameter of 1.5 mm, which we soldered in a loop around the growth chamber (not shown in Figure 2). The pumping conductance of this tube is quite small (approximately 0.01 Torr-liter/second near operating pressure), and its temperature is equal to that of the growth chamber. The remainder of the pumpout tube is not the same temperature as the growth chamber, but the low conductance means the outer part of the tube cannot contribute significantly to $\sigma_{surf}$ around the sample crystal. Note that the chamber volume was approximately 0.07 liters, so the pumpout speed remained high enough to pump out the chamber in tens of seconds. We typically pumped the chamber out at a low rate, about 2-3 Torr/second at most, to avoid evaporating the sample crystal.

Pumping on the chamber removes water vapor from the ice reservoir, and this causes its surface to cool. Our calculations showed that this cooling could be significant in that it would affect our determination of $\sigma_{surf}$. We reduced this problem simply by not pumping on the chamber when acquiring growth data. Once pumping is stopped, the timescale for the ice reservoir to reach temperature equilibrium with the chamber is

$$\tau \approx \frac{C\rho L^2}{\kappa} \tag{5}$$

where $C \approx 2000$ J/kg-K is the heat capacity of ice, $\rho \approx 917$ kg/m$^3$ is the density of ice, $L \approx 3$ mm is the thickness of the ice, and $\kappa \approx 2.4$ W/m-K is the thermal conductivity of ice. Plugging these numbers in gives a relaxation time of less than 10 seconds. We typically allowed at least a minute after pumping for the ice reservoir and the sample crystal to reach a steady state.

**Temperature Gradients in the Transfer Tube.** The transfer tube extends from the growth chamber into the large tank (shown in Figure 2), and thus there may be a temperature gradient along the tube. This gradient is large when the growth chamber is at a significantly different temperature from the large tank. Making the conductance of this tube small (like with the pumpout tube) was not an option, as this would adversely affect the transfer efficiency.

To reduce the systematic errors associated with the temperature profile of the transfer tube, we installed an additional valve at the growth chamber (V2 in Figure 2). This is a teflon-in-copper ball valve that does not form a very tight vacuum seal; the task of sealing is relegated to the valve V1. The additional valve does have a low conductance in the off position, however, and the body is made of copper soldered directly to the growth chamber, so the two have the same temperature. When V2 is in the off position, the low conductance means the remainder of the transfer tube is isolated from the growth chamber. We close V2 once a sample crystal is in place on the substrate.

**Temperature Equilibration of the Growth Chamber.** As mentioned above, the crystal transfer from the nucleation chamber to the growth chamber must take place when both are at nearly the same temperature. After the transfer, we then sometimes heat or cool the growth chamber to reach some target temperature for a measurement. The timescale for equilibration of the chamber is again given by Equation 5, except using quantities for copper. Taking $C \approx 400$ J/kg-K, $\rho \approx 9000$ kg/m$^3$, $\kappa \approx 400$ W/m-K, and $L \approx 5$ cm gives $\tau \approx 20$ seconds. Changing the chamber temperature takes several minutes, while monitoring the sample crystal to make sure it does not grow or evaporate significantly during the change. Because the timescale for change is slow, we believe the errors resulting from a lack of temperature equilibration are negligible.

**Temperature Equilibration of the Substrate.** During a typical growth run, we first let the sample crystal come into steady state with the rest of the chamber (at $\Delta T \approx 0$). Then we slowly increase the substrate temperature until the crystal begins to evaporate, which establishes a $\sigma_{surf} = 0$ calibration point. We then increase $\sigma_{surf}$ in jumps, stopping after each jump to count fringes and thereby generate a growth velocity measurement. The video is recorded to DVD during this process, so that $\Delta T$, $\sigma_{surf}$, and $v_n$ can all be determined later.

When increasing $\sigma_{surf}$, we determine the substrate temperature using a small thermistor located in the copper base just below the substrate (see Figure 4). We assume that the temperature at this thermistor is equal to the temperature of the substrate, and thus the temperature of the sample crystal. After a temperature jump, the equilibration time for the substrate is given by Equation 5, this time using quantities appropriate for sapphire. With $C \approx 700$ J/kg-K, $\rho \approx 4000$, $\kappa \approx 40$ W/m-K, and $L \approx 2$ mm, we have $\tau \approx 0.3$ seconds, which is much faster than our measurement process. We note that glass has a thermal conductivity that is nearly 40 times lower than sapphire, giving $\tau \approx 10$ seconds in that case, which is comparable to our measurement time.

**Heating Effects.** The latent heat generated by the crystal growth is readily transferred to the substrate, with only minor heating of the growing crystal [1]. Thus heating effects are negligible in our measurements.

**Chemical Contamination.** This is always a wild card in our experiments, since we do not yet know how clean is clean enough for our apparatus. The air in the nucleation chamber is ordinary laboratory air, so we have the possibility that the growing ice crystals become coated with chemical contaminants that affect their growth. We have observed, however, that crystals grown in laboratory air clearly show the growth characteristics described by the morphology diagram [1]. This gives us reason to believe that these contaminants are not greatly affecting the crystal growth rates.

We have several observations that suggest that contamination is playing a minor role in our measurements. First, we often see small regions on crystal surfaces that do not fill in to form flat facets; Figure 2 shows one example of this. We believe that growth in these regions is prevented because of a buildup of impurities in those spots (although the evidence for this is not conclusive). As the crystal grows, the slow-growing region is left behind as the remainder of the crystal grows relatively free of contaminants.

Second, we sometimes observe that the supersaturation must be abnormally high before a crystal starts growing, as if the growth has to break through a "shell" of impurities. To get around this effect (whatever its cause), we first cool a sample crystal until it grows and let it increase in thickness by 2-3 microns. This seems to be enough to produce fresh ice with growth that is largely unaffected by impurities.

Third, we found that impurities can be picked up from the substrate as well as from the air. Figure 6 shows a crystal growing on a substrate that was coated with solvent residue. The residue affected the crystal growth substantially. After noticing these substrate coating problems, we subsequently cleaned the substrate only with deionized water, which is done before every run.

We made a substantial effort to keep solvents and other volatile materials out of our growth chamber, shown in Figure 4. In particular, the thermoelectric module, which needs some sort of grease to make a good thermal contact, lies outside the vacuum chamber. The chamber is predominantly copper, and we used a small amount of thermally conducting epoxy to hold the sapphire window to its copper base. We also used some vacuum compatible grease on an o-ring that seals the cover to the chamber. We also gently bake the chamber regularly in air to remove residual solvents and high vapor pressure materials. After cooling the chamber before a growth run, we cycle the air several times, replacing it with cold air inside the nucleation chamber.

**Temperature Drifts.** A high degree of temperature stability was necessary for obtaining satisfactory growth data. The temperature of the growth chamber is held constant to better than 0.03 C for several hours during a run (and is known with an absolute accuracy of approximately 0.1 C). The temperature of the substrate is determined to 0.01 C relative to the chamber, as measured by a differential temperature controller.

**Crystal-to-Crystal Variations.** Even with a perfect experimental apparatus, there are still substantial crystal-to-crystal variations in growth. We have found it absolutely necessary to examine many crystals, and to use at least ten to produce an accurate picture of $\alpha(\sigma)$ for each facet. Dislocations are certainly a factor, and we occasionally encounter "fast-growing" crystals that grow abnormally rapidly at low $\sigma_{surf}$. We have also found that evaporating a large crystal and regrowing it leads to increased dislocations and perhaps impurity problems. For best results, it is necessary to use a new crystal for each growth measurement.

**Summary.** After observing many growing crystals, we have found that the following were important for producing reliable growth data:

1) We needed to grow small crystals in low pressure. This was necessary to avoid diffusion limiting effects, which are remarkably difficult to model accurately.

2) We needed to look only at the growth of facets not in contact with the substrate. Interaction with the substrate may (or may not, depending on poorly understood physics) substantially affect the growth behavior of ice crystals.

3) The sample crystal could not have any neighbors on the substrate. We made sure the entire substrate surface was viewable, and we used a $CO_2$ laser to remove neighbor crystals. Until we controlled the neighbor problems, our data showed large variations in growth rates.

4) We needed excellent temperature control of our apparatus, controlling important surfaces to tens of millidegrees. A sapphire substrate was used to reduce temperature errors between the servo sensor, the substrate surface, and the growing crystal.

5) The growth chamber needed to be highly uniform in temperature, except for the substrate. In particular, all ice inside the chamber (except for the sample crystal) needed to be at the same temperature. A large ice reservoir helped stabilize $\sigma_{surf}$ as well.

6) We needed to grow many crystals. This was useful for observing and controlling systematics, but we also found that not all crystals grow the same way. The presence of dislocations and/or contaminants may produce different growth behaviors. This is true even if one chooses only crystals with clean, prism-like morphologies.

## 5. Initial Results

To test our measurement apparatus and techniques, we made a number of measurements of the growth of the basal facets of ice crystals at a temperature of -15 C. The nucleation chamber and growth chamber were kept at this temperature, and neither was changed during the measurements. A typical data run went as follows:

1) Cool the nucleation tank and growth chamber until both are stable (typically 4-6 hours).

2) Add water to the vessel in the nucleation tank and let the temperature stabilize (one additional hour). Start the nucleator to produce ice crystals.

3) Begin transferring crystals from the nucleation tank into the growth chamber. Let the transferred crystals evaporate on the warmed substrate until a suitable sample is obtained, then cool the substrate so the crystal neither grows nor evaporates. A good sample will have a clean prism morphology and a size between 20 and 50 microns. The basal facets should be parallel to the substrate so the interferometer produces good fringes.

4) Use the $CO_2$ laser to evaporate away any neighboring crystals.

5) Slowly pump the air out of the growth chamber so the pressure goes down at 2-3 Torr per second. Adjust the substrate temperature during this process so the sample crystal remains stable. Stop pumping when the pressure is below 3 Torr and let the system stabilize for a few minutes.

6) Heat the substrate until the sample crystal begins to evaporate. Estimate the substrate temperature servo set point at which $\sigma_{surf} = 0$ to an accuracy of 0.01 C.

7) Cool the substrate until the crystal begins growing. Let the crystal grow thicker by 2 microns (about 10 fringes of the interferometer) before making any measurements.

8) Warm the substrate until the crystal stops growing and then set the temperature for slow growth. Continuously record the temperature setting and the video signal to DVD.
9) Cool the substrate again, let it stabilize for several seconds, so that the interferometer fringes record the growth. Repeat this process of cooling the substrate in jumps followed by a growth measurement.
10) Check for hysteresis by warming the crystal again so the growth slows, and again record the growth.
11) At end of crystal run, heat the substrate to remove all ice. Bring the pressure back to one atmosphere.
12) Go back to step (3) and start with another crystal.
13) After collecting data for several hours, allow the apparatus to warm up to ambient temperature. Clean the substrate as necessary before another run.
14) After the run, transcribe the data on the DVD for each crystal, to determine the growth velocity $v_n$ and supersaturation $\sigma_{surf}$ for each temperature point. The end result is a plot of $v_n(\sigma_{surf})$, or equivalently $\alpha(\sigma_{surf})$, consisting of several points for each crystal.

    Figures 8 and 9 show data taken on two separate days of running. The apparatus was warmed, and the substrate was cleaned, between these runs. A number of crystals were rejected before the data were transcribed from DVD, and these are not shown in the plots. One reason for rejection was if the basal facet was not parallel to the substrate, so laser fringes could not be discerned in the video images. Another reason for rejection was if neighbor crystals were seen at the end of the data taking for a particular crystal. Often the neighbor crystals did not appear until after growth data were being taken, so it was too late to remove them using the $CO_2$ laser.

    Two of these crystals were sufficiently different from the rest that we chose to remove them before merging the data, although they are shown in Figures 8 and 9. Crystal 8 on 6/4/06 was a clear "fast-grower", showing rapid growth at low $\sigma$ from the beginning. The data show approximately $v_n \sim \sigma^2$, or equivalently $\alpha \sim \sigma$, consistent with growth driven by a spiral dislocation on the basal surface. Crystal 2 on that same day grew substantially slower than average, for unknown reasons. We suspect the growth may have been slowed by an unseen neighbor crystal, but this is not known for certain.

    We also found that some crystals showed abnormally fast growth at low $\sigma$ at the beginning of a run, for example Crystal 5 on 6/4/06. We are still investigating this behavior and trying to understand its origin. At the risk of distorting our results, we removed the first few data points from Crystals 5 and 6 on 5/31/06, and from Crystals 5 and 10 on 6/4/06. We also removed the last two points from Crystal 7 on 5/31/06, as these points were taken when the crystal was large and $\sigma$ was large, and under such conditions we believe the growth measurements are not accurate. Finally, we removed the last three points from Crystal 6 on 6/4/06, owing to neighbor issues. Figure 7 shows a view of the substrate after the data for Crystal 6 were taken, showing the emergence of two neighbors. These neighbors are small, far away, and appeared late, so we believe they only significantly affected the last few points.

    Although purists may balk at our removal of suspicious data points, we believe it at least somewhat justified by our experience watching numerous growing crystals. In particular, the data would be badly skewed if crystals like Crystal 8 were not thrown out, since the growth in these abnormal cases is clearly dominated by

different physical mechanisms than is the norm. Removing the additional points is essentially a form of "robust" fitting, where a few percent of the outlier points are removed so they do not adversely contaminate an otherwise sound data set. Here again, we believe that are not just outlier points, but that the growth is being influenced by physical mechanisms that are not usually present in ice crystal growth. To appease the purists, we note that Figures 8 and 9 do include all the data taken during these runs.

The remaining data points from these two days of measurements were combined to produce the plots in Figure 10. The data are consistent with growth in which the attachment kinetics are dominated by 2D nucleation at the facet surface. Further interpretation of these data will require additional experiments at different temperatures and additionally looking at the growth of prism facets. We are currently working on such a set of measurements.

## 6. Comparison with Previous Experiments

In our recent review of ice crystal growth data, we found that essentially all previous experiments produced unreliable data. With the new results presented above, we must now add our own previous experiment [4] to the slag heap. Upon close examination of our prior experiment, we found that a number of systematic errors had not been adequately dealt with. In the first place, we did not appreciate how greatly neighbor crystals reduced $\sigma_{surf}$ around our sample crystals. We also used a glass substrate, which had a rather slow equilibration time, as discussed above. Even worse, the substrate was not all visible, so unseen neighbors probably played some role in reducing the measured growth rates. Finally, our growth chamber did not have a sufficiently uniform temperature distribution, which again likely affected our data. As a result, our current data at -15 C are much more trustworthy and show much higher growth rates.

The potential for additional systematic errors still exists, and we will continue to investigate these possible problems as we push our experiments to other temperatures and conditions. We believe we have made much progress in understanding and eliminating these persistent problems, however, and that our current data are, for the first time, giving an accurate picture of the growth rates of faceted ice crystals.

# 8. Figures

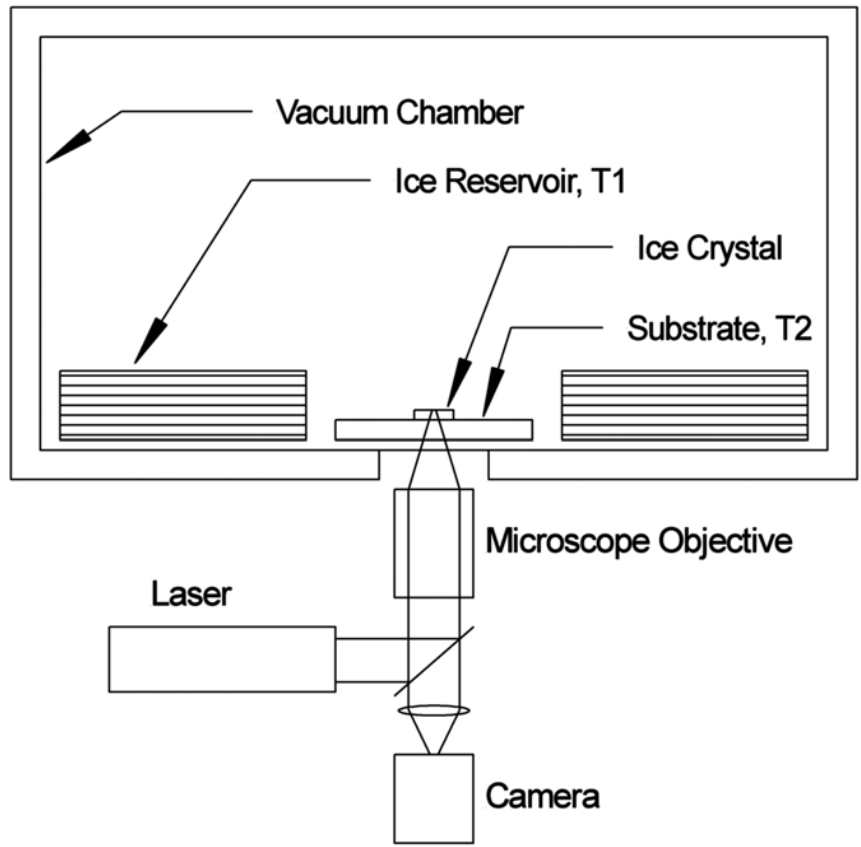

Figure 1. The basic layout of our experimental apparatus. An ice crystal sample is placed with known orientation on a substrate inside an evacuated growth chamber. An ice reservoir inside the chamber provides a source of water vapor to grow the sample crystal. The supersaturation is determined by the temperature difference between the ice reservoir (equal to the temperature of the rest of the growth chamber) and the substrate. The sample crystal is imaged using a microscope objective and a video camera. A low-power laser is focused onto the crystal by the same microscope objective. The laser spot is reflected by the top and bottom of the ice crystal, and the two reflections interfere. The brightness of the reflected laser spot, as seen in the camera, thus cycles as the crystal grows thicker.

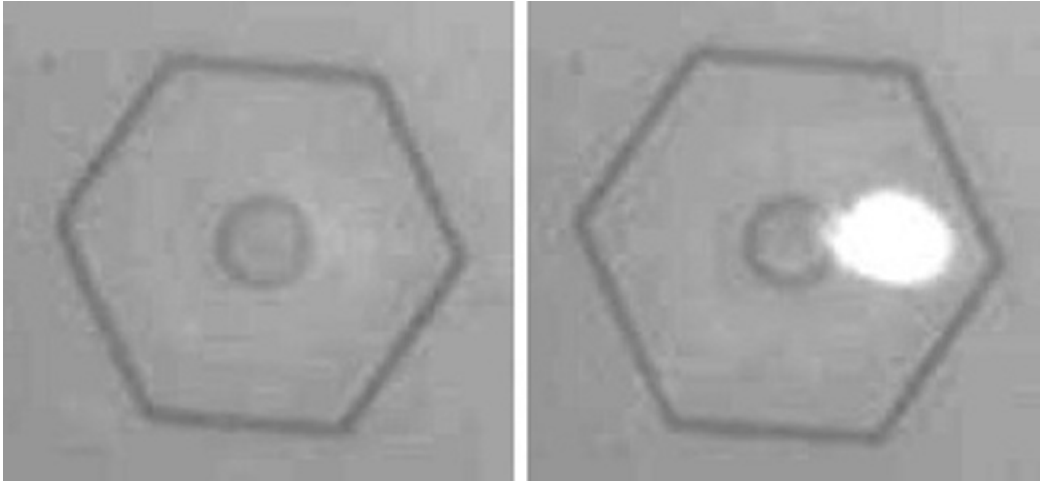

Figure 2. Image of a typical crystal growing on the substrate. The crystal has a hexagonal prsim morphlogy, about 50 $\mu$m in diameter, with one of the basal faces lying flat on the substrate. The picture on the left shows the crystal when the two laser reflections (described in the text) were interfering destructively. The picture on the right was taken about 20 seconds later, when the crystal had grown and the two laser reflections were interfering constructively. The central spot is a depression in the crystal that is not filling in, probably because this is a region of concentrated impurities on the crystal surface. (This is discussed further in the section on systematic errors.)

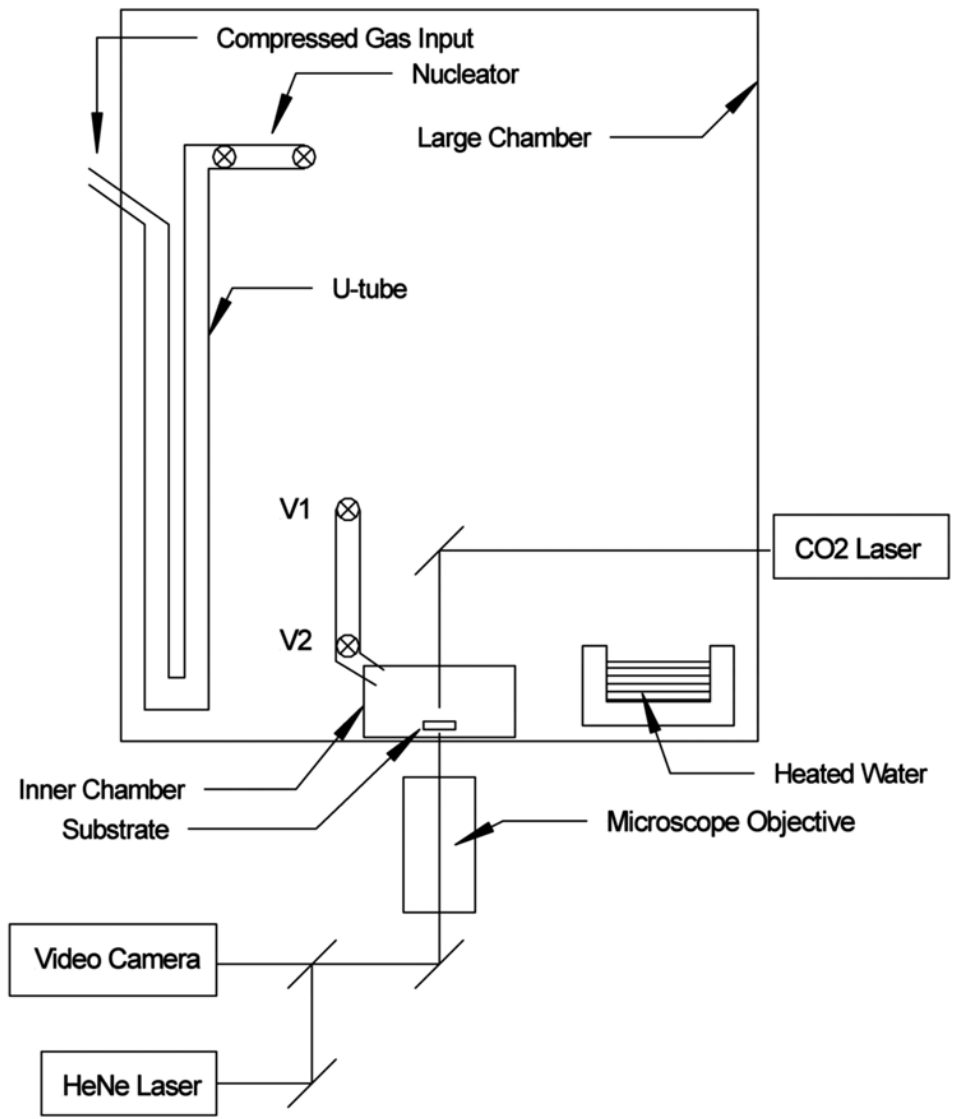

Figure 3. A schematic diagram of our nucleation chamber for producing small ice crystals. The crystals are created in air, and they grow until they begin to fall under gravity. Drawing air from the nucleation chamber to the smaller growth chamber transfers some ice crystals through valves V1 and V2, and by chance some crystals land on the substrate inside the growth chamber. After the crystal transfer, the valves are closed to isolate the growth chamber, which is then evacuated so growth measurements can be made.
A schematic diagram of the ice crystal growth chamber. Care was taken to ensure that the chamber temperature was uniform except for the sapphire substrate, which was set using a differential temperature controller.

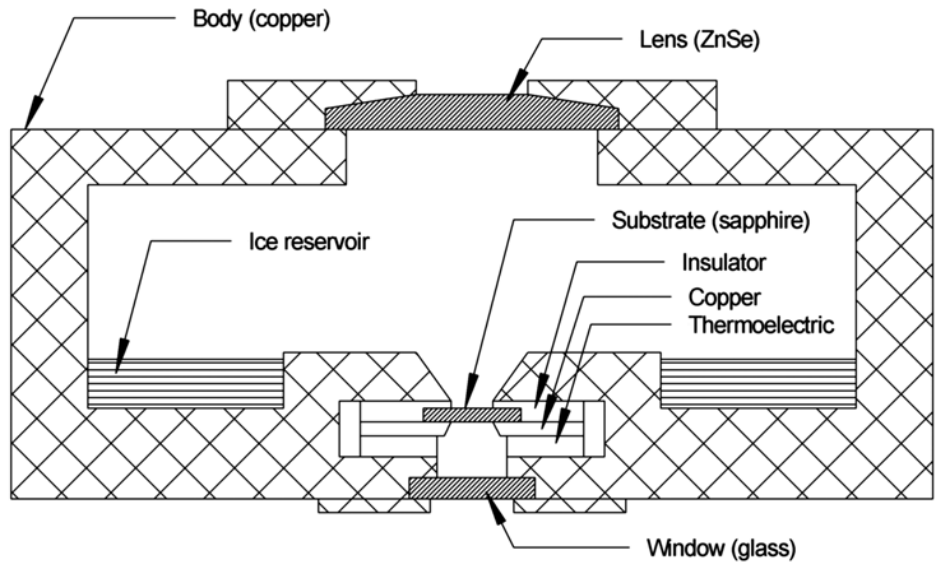

Figure 4. A schematic diagram of the ice crystal growth chamber. Care was taken to ensure that the chamber temperature was uniform except for the sapphire substrate, which was set using a differential temperature controller.

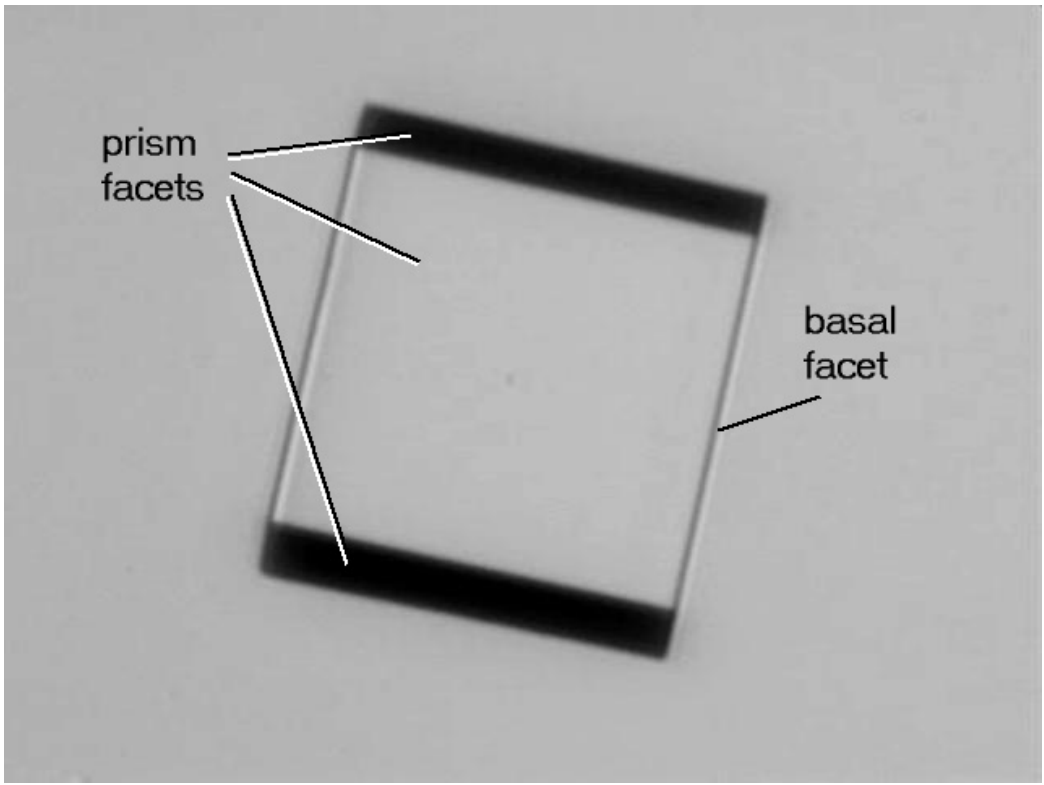

Figure 5. This picture shows an ice crystal after a period of extensive growth on the substrate. The overall size of the crystal is about 240 $\mu$m. Its morphology was initially that of a simple hexagonal prism with one prism facet lying flat against the substrate. As the crystal grew, the prism facets intersecting the substrate grew much more rapidly than the prism facet that did not touch the substrate, giving the crystal the shape seen here. This clearly demonstrates that the facet growth can be substantially affected by interaction with the substrate.

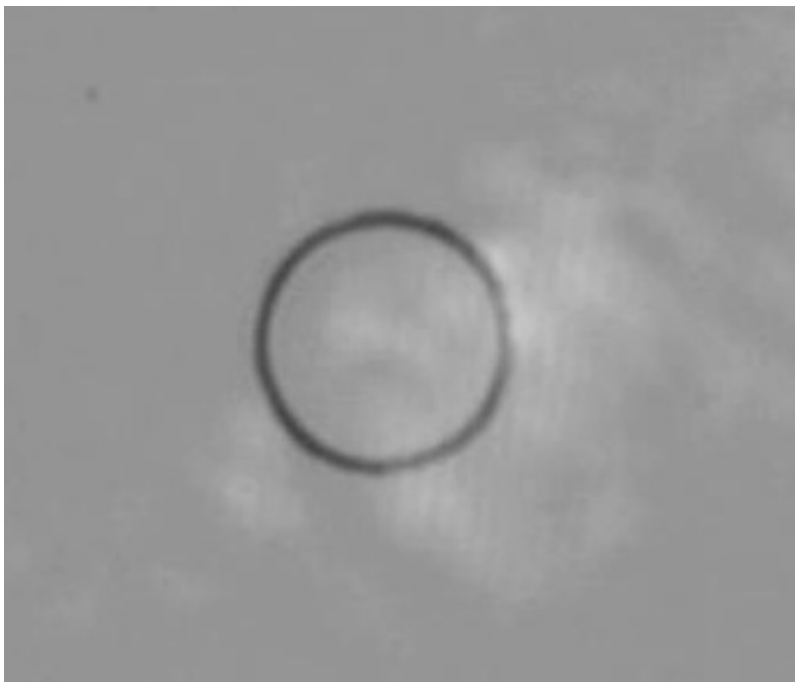

Figure 6. This picture shows a plate-like crystal with no prism facets. It grew on a substrate that was contaminated with a solvent residue from cleaning. The residue was apparently picked up by the crystal, where it prevented the growth of prism facets. This unusual growth behavior is not present when the substrate is cleaned with distilled water.

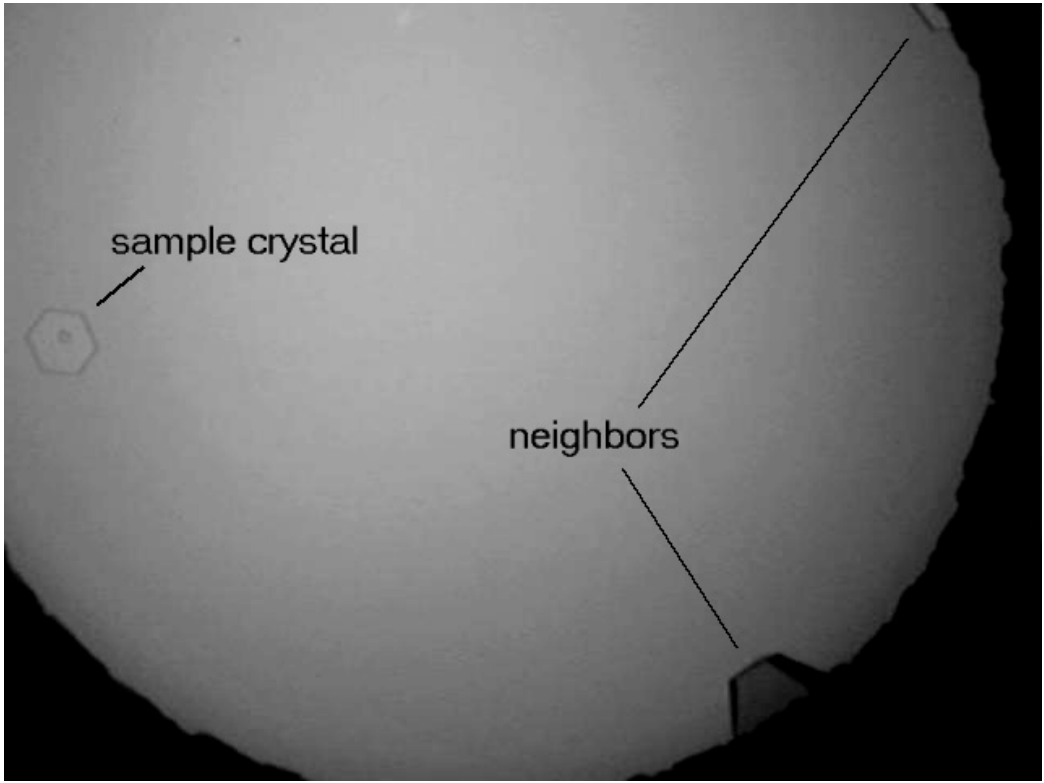

Figure 7. A view of the substrate after growing Crystal 6 on 6/4/2006. Several neighbors were seen growing far from the sample crystal. These neighbors were small enough, and far enough away, that they apparently did not greatly affect the growth of the sample crystal. The clear diameter of the substrate in this view is three millimeters.

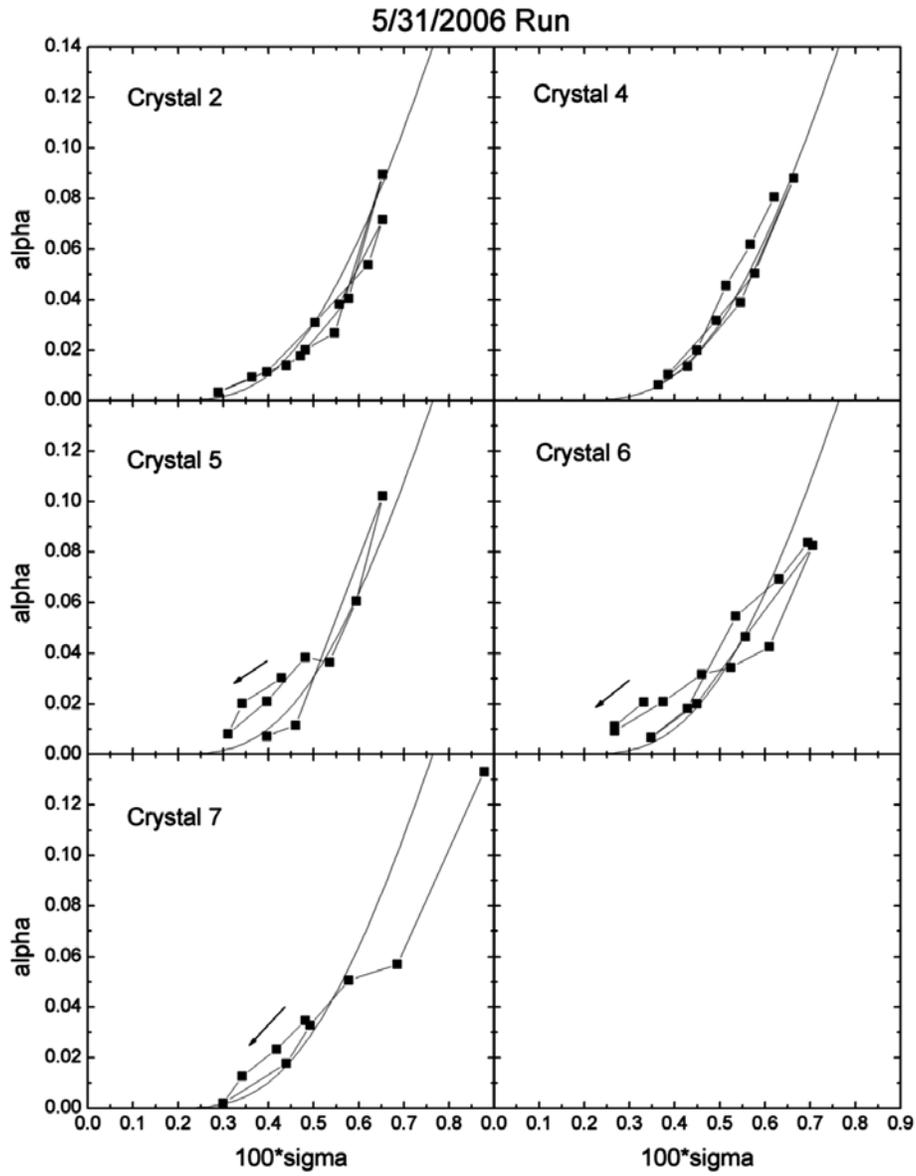

Figure 8. Data collected during a one-day run. Several crystals were rejected before the data were transcribed (see text), and these are not shown. Arrows show the direction in which points were taken, usually from low $\sigma$ to high $\sigma$, and then back again. The smooth curve shows $\alpha(\sigma) = 2\exp(-0.021/\sigma)$ in all plots.

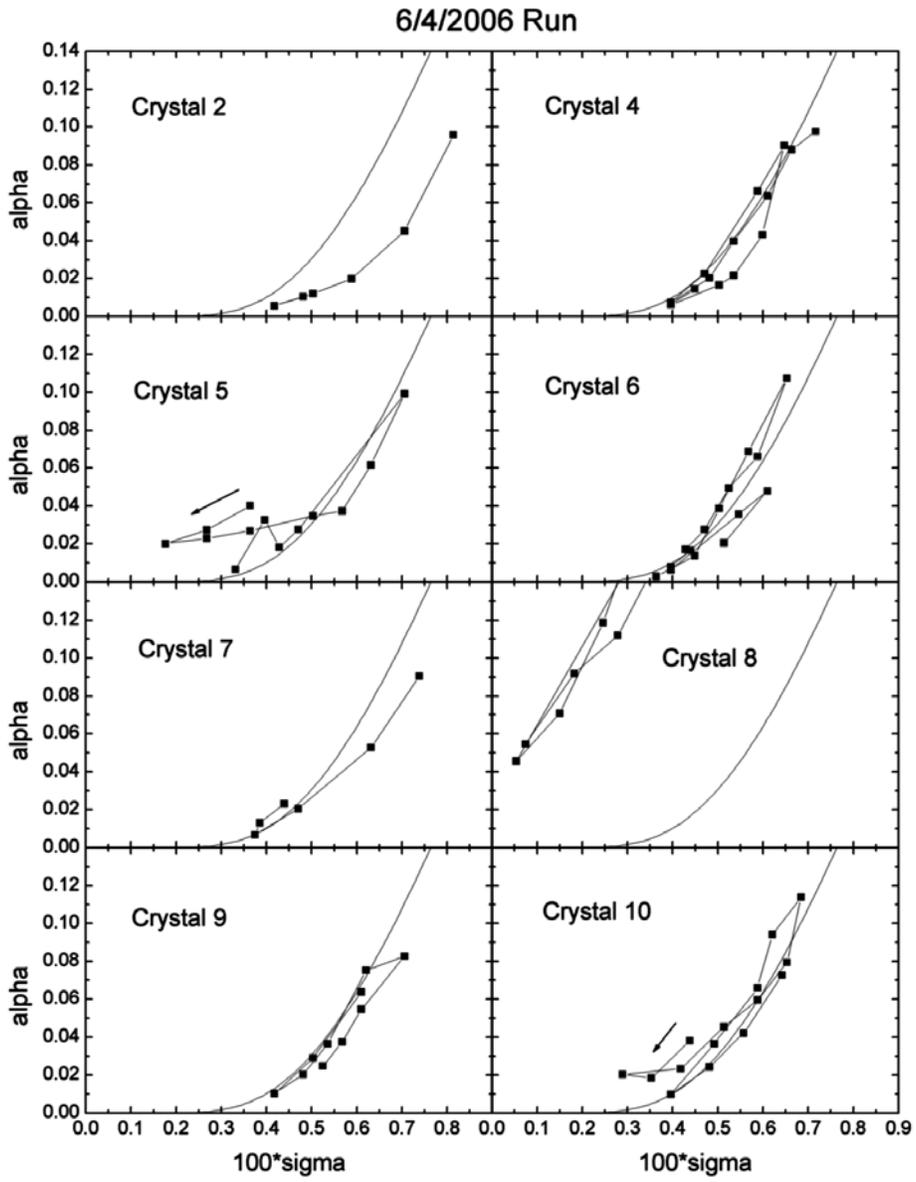

Figure 9. Same as the previous figure, but from a second day of data taking.

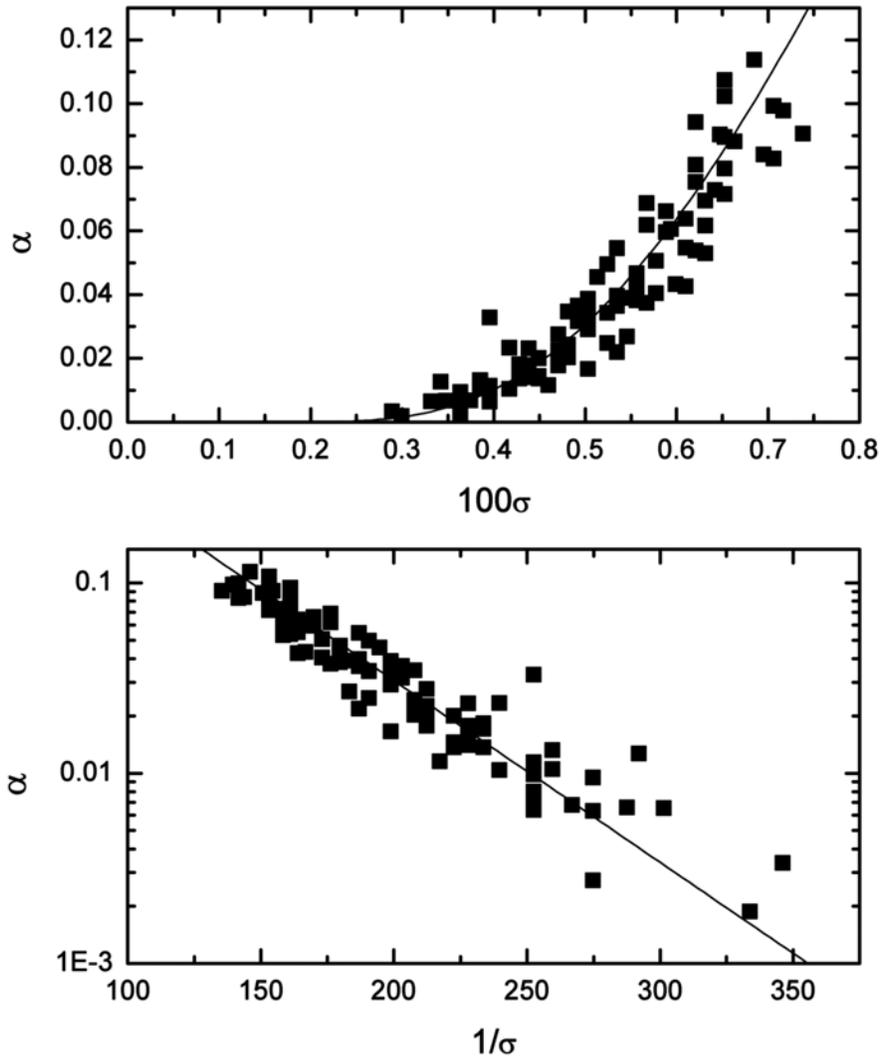

Figure 10. Combined data from both runs, after removing data as described in the text. The two plots show the same data plotted different ways. The curve, $\alpha(\sigma) = 2\exp(-0.021/\sigma)$ has the functional form expected for nucleation-limited growth.